\documentclass[sigconf]{acmart}
\AtBeginDocument{%
  }

\usepackage{amsmath}
\ifPDFTeX
  \usepackage{bm}
\else
  \newcommand{\bm}[1]{\symbf{#1}}
\fi
\usepackage{multirow}

\setcopyright{none}
\acmConference[WWW '27]{The ACM Web Conference 2027}{May 10--14, 2027}{Dublin, Ireland}
\acmYear{2027}
\copyrightyear{2027}
\acmDOI{}
\acmISBN{}
\author{Ziyang Su}
\affiliation{%
  \institution{Shanghai Jiao Tong University}
  \city{Shanghai}
  \country{China}}
\email{suziyang@sjtu.edu.cn}

\author{Guanjie Zheng}
\affiliation{%
  \institution{Shanghai Jiao Tong University}
  \city{Shanghai}
  \country{China}}
\email{gjzheng@sjtu.edu.cn}

\author{Jilin Hu}
\affiliation{%
  \institution{East China Normal University}
  \city{Shanghai}
  \country{China}}
\email{hujilin1229@gmail.com}

\author{Linghe Kong}
\affiliation{%
  \institution{Shanghai Jiao Tong University}
  \city{Shanghai}
  \country{China}}
\email{linghe.kong@sjtu.edu.cn}

\newcommand{\model}{\textsc{DualClusterNet}}
\newcommand{\R}{\mathbb{R}}
\newcommand{\concat}{\mathbin{\|}}
\newcommand{\sigmoid}{\operatorname{sigmoid}}
\newcommand{\softmax}{\operatorname{softmax}}
\newcommand{\diag}{\operatorname{diag}}
\newcommand{\LN}{\operatorname{LayerNorm}}

\title[DualClusterNet for Next-POI Recommendation]{DualClusterNet: Personalized Pattern-Space Composition for Next-POI Recommendation}

\begin{document}
\begin{abstract}
Next point-of-interest (POI) recommendation supports personalized place discovery on location-based Web platforms. Mobility routines can recur across users while being realized at different places and adapted to individual preferences. We present \model, a causal sequential recommender that composes two population-shared prototype dictionaries. At each observed prefix, soft routers retrieve pattern and spatial mixtures; a factorized multiplicative interaction couples them, feature-wise affine modulation adapts the shared representation, and a separate user-conditioned branch with an additive--projective gate retains individual variation. The prediction head combines this representation with a learned first-order transition term and ranks the complete training POI vocabulary. We evaluate on NYC and TKY using a training-only vocabulary and an original-final-target protocol that neither bridges unknown POIs nor substitutes earlier targets. The available single-seed results show competitive full-candidate ranking, while controlled analyses support the contribution of the compositional and personalized branches. Code is available at \url{https://anonymous.4open.science/r/DualClusterNet-F4EE/}.
\end{abstract}

\maketitle

\section{Introduction}
\label{sec:introduction}

Location-based Web platforms connect online place discovery with users' physical activities. Their ranking systems must select useful destinations from a city-wide catalog while accounting for the order, timing, and geography of past visits. Next point-of-interest (POI) recommendation formalizes one part of this problem: given an observed check-in history, rank the concrete POI that the user will visit next~\cite{stan,getnext}. This work addresses the \emph{Search, Recommendation, and Retrieval-Augmented AI} setting through a specific Web recommendation challenge: reusing behavioral regularities across users whose spatial contexts and individual preferences differ. The prediction concerns the immediate next place, not a route, a new-to-the-user destination, or a POI conditional on a supplied future timestamp.

A useful distinction is between a routine and its realization. Consider an illustrative routine such as visiting a workplace, obtaining a meal, and then stopping at a shop. Two users may exhibit a similar ordering of activities without sharing a single POI. Conversely, the same neighborhood may host different routines at different times, and two users in the same neighborhood need not prefer the same restaurant. A model should therefore have a way to reuse population-level regularities, condition them on where a trajectory is unfolding, and retain user-specific variation. The example motivates an inductive bias; it does not presume that latent clusters are already validated semantic labels such as ``commuting'' or ``shopping.''

Existing methods provide complementary ingredients. Recurrent and attention-based models encode temporal and spatial dependencies~\cite{strnn,stgn,stan}. Long- and short-term preference methods distinguish persistent tastes from recent behavior~\cite{plspl}. Graph and hypergraph approaches exploit transitions and collaborative relations across users~\cite{getnext,sthgcn,dchl}. Recent work also explicitly models latent preferences or personalized spatio-temporal clusters~\cite{cllp}. Consequently, neither sharing information across users nor introducing clusters is, by itself, the distinction of this work. Our focus is the \emph{composition rule}: how two reusable dictionaries interact at each trajectory prefix, and how their joint representation is adapted to a user before ranking places.

This focus yields three design questions. First, a trajectory may move between routines rather than belonging permanently to one cluster. The assignment should therefore depend on its current prefix and remain soft. Second, spatial context should affect the expression of a pattern, rather than simply contribute an unrelated feature to a concatenation. Third, shared structure should not force users to have identical representations. An additive residual and a user-conditioned interaction offer different ways to preserve individuality, and their relative influence can depend on the current state.

We instantiate these ideas in \model. A causal LSTM encodes the observed POI, category, temporal, geographic, and user context. Two assignment networks retrieve mixtures from a pattern dictionary and a spatial dictionary. An elementwise product of their projected mixtures implements a factorized interaction over all pattern--space pairs. Feature-wise affine modulation then adapts this shared interaction to the user. A separate user-conditioned branch captures information not constrained to pass through the dictionaries. Finally, a scalar gate mixes additive fusion with an affine transform of the branches' multiplicative interaction. A full-candidate score combines the resulting preference representation with a learned first-order POI transition term.

Evaluation must preserve what ``next'' means. If an unseen POI interrupts a trajectory, deleting it and joining the remaining visits introduces a transition that was not observed. If the final POI is unknown, replacing it with an earlier known POI changes the prediction target. We instead use a training-only candidate vocabulary and retain only the final continuous known suffix for each held-out trajectory. All compared implementations receive the same valid queries, candidate set, ranking metrics, and validation-based checkpoint selection rule. This defines a closed-set experiment; it is not an evaluation of unseen-user or unseen-POI generalization.

The contributions of this study are as follows.
\begin{itemize}
    \item We introduce two shared dictionaries whose assignments adapt to the observed prefix, together with a pattern--space interaction and separate personalized branches.
    \item We give an explicit factorization of its pairwise interaction and an additive--projective fusion rule, making the roles and limits of the shared and personalized components precise.
    \item We provide a unified-protocol comparison with nine local baseline implementations on two cities. The current single-seed results lead NYC Acc@1 and MRR and remain competitive on TKY; they do not establish uniform superiority across metrics or cities.
\end{itemize}

\section{Related Work}
\label{sec:related}

\subsection{Sequential and Context-Aware POI Recommendation}

Sequential recommendation combines personal preference with dependence on recent observations. FPMC factorizes personalized Markov transitions~\cite{fpmc}; we adapt its singleton-target form for next-POI scoring. ST-RNN adds spatial and temporal context to recurrent transitions~\cite{strnn}. STGN and its coupled-gate STGCN variant control recurrent updates with spatio-temporal inputs~\cite{stgn}. PLSPL combines long- and short-term preferences~\cite{plspl}. STAN uses spatio-temporal attention beyond adjacent check-ins~\cite{stan}. DeepMove attends over historical trajectories to capture mobility periodicity~\cite{deepmove}, while Flashback retrieves past recurrent states according to spatio-temporal similarity~\cite{flashback}. LSTPM pairs a nonlocal long-term preference module with a geography-aware short-term state~\cite{lstpm}. Together, these methods show that useful context may come from distant history as well as the latest transition.

Historical attention and prototype composition represent different sources of reuse. Attention and state retrieval reweight representations tied to recorded visits, whereas our dictionaries are population-shared parameters that can be mixed even when users do not share POIs. Likewise, long- and short-term modules separate temporal horizons, while \model\ separates pattern and spatial axes and lets the spatial mixture modulate the pattern mixture. It uses a single causal recurrent encoder, does not require a supplied next-visit timestamp, and ranks the complete candidate vocabulary. This distinction motivates evaluating the interaction itself rather than treating memory, attention, and clustering as interchangeable mechanisms.

\subsection{Collaborative, Graph, and Hypergraph Models}

GETNext adds a global trajectory flow map to a Transformer~\cite{getnext}. STHGCN captures higher-order trajectory relations with spatio-temporal hypergraphs~\cite{sthgcn}. DCHL learns disentangled collaborative, transition, and geographic views~\cite{dchl}. HVGAE studies hyperbolic graph representations and variational next-POI modeling~\cite{hvgae}. These methods expose population-level information beyond one user's sequence, but construct their relations and representations differently.

Our model does not perform graph message passing or hypergraph convolution. It shares prototype parameters across the training population and uses a prefix-conditioned combination of them. This is a different architectural bias, not a claim that dictionary sharing is a substitute for every form of collaborative graph information. In particular, the dictionaries are learned jointly with the recommendation objective and are not extracted as communities from a user--POI graph.

\subsection{Latent Preferences, Clustering, and Personalization}

Clustering and multi-interest modeling are closely related to our motivation. CLLP models latent preferences using preference prototypes, distillation, and contrastive learning~\cite{cllp}. DPRL investigates disentangled and personalized representations for next-POI recommendation~\cite{dprl}. These are important neighboring approaches: the use of a latent pattern bank or personalized spatial information is already established, so a novelty claim must concern the particular interaction and conditioning mechanism rather than clustering alone.

\model\ separates two dictionary axes and combines their mixtures multiplicatively before user-specific modulation. The modulation follows the feature-wise affine conditioning principle of FiLM~\cite{film}, with sigmoid-constrained scales in our implementation. A second user-conditioned branch and a learned gate distinguish additive correction from multiplicative interaction. Unlike objectives designed to enforce independence between latent factors, our next-POI loss contains no decorrelation, contrastive, or cluster-balancing term. Accordingly, ``shared'' denotes shared parameters and reusable components, not an identified user-invariant subspace. Section~\ref{sec:ablation} tests the predictive contribution of the composition and personalization components; semantic interpretability of the learned factors remains an open question.

\section{Methodology}
\label{sec:method}

\begin{figure*}[t]
    \centering
    \includegraphics[width=\textwidth]{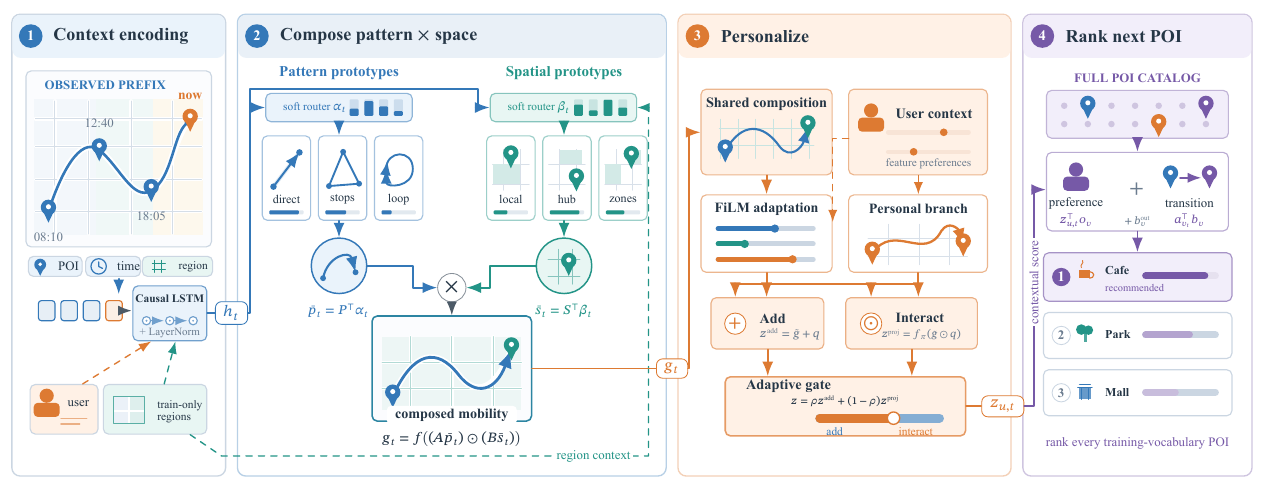}
    \caption{Overview of \model\ at an observed trajectory prefix. (1) A causal encoder uses only observed check-ins and training-derived context. (2) Prefix-dependent soft routers mix population-shared pattern and spatial prototypes; their projected mixtures interact multiplicatively to express a mobility pattern in its current spatial context. (3) User FiLM and a separate personal branch feed additive and projective paths, which an adaptive scalar gate interpolates. (4) Contextual preference, the transition from the latest observed POI, and candidate bias rank the full training POI vocabulary. Prototype-card labels are schematic motifs rather than learned semantic assignments. Training-only geographic regions are context features and are distinct from the learned spatial prototypes.}
    \Description{Four-stage architecture illustrated with trajectories and maps. Observed check-ins enter a causal encoder. Soft routing mixes reusable trajectory-pattern cards and spatial-map cards, then overlays the resulting route and spatial context through multiplicative composition. User feature sliders and a separate personal route feed additive and interaction paths combined by an adaptive gate. Contextual and latest-POI transition signals score the full catalog to produce a ranked next-POI list.}
    \label{fig:framework}
\end{figure*}

Figure~\ref{fig:framework} summarizes the DualClusterNet computation. The architecture comprises a causal context encoder, pattern--space composition, user-specific adaptation, and a full-candidate prediction head.

\subsection{Problem Formulation and Information Boundary}
\label{sec:problem}

Let $\mathcal U$, $\mathcal V$, and $\mathcal C$ denote the user, POI, and category vocabularies observed in the training split. Each POI $v\in\mathcal V$ has training-derived coordinates $\ell_v$ and category $c_v$. A user's observed trajectory prefix is $H_{u,t}=(x_1,\ldots,x_t)$, where $x_j=(v_j,\tau_j)$ records a POI and timestamp. The task is to rank every candidate $v\in\mathcal V$ according to a score $s(v\mid u,H_{u,t})$, with the actual next POI $v_{t+1}$ as the single target. Repeated visits are valid targets. The future timestamp $\tau_{t+1}$, future location, and future category are not query inputs.

We learn parameters from training trajectories and select a checkpoint on validation trajectories. For held-out evaluation, an unknown POI terminates the current known segment. Only the continuous known suffix ending at the original final check-in is eligible, and it must contain at least one historical check-in plus its target. An unknown user or final target makes the trajectory ineligible. This policy is formalized in Section~\ref{sec:protocol}. It preserves the original final prediction question rather than inventing a new transition after filtering.

We use $d_e$ for the POI embedding dimension, $d_c$ for each context embedding dimension, and $d$ for the hidden dimension. The number of geographic partitions is $R$, the number of pattern prototypes is $K_p$, and the number of latent spatial prototypes is $K_s$. These three counts need not coincide.

\subsection{Geographic Context and Causal Trajectory Encoding}
\label{sec:encoder}

\paragraph{Training-only geographic partitions.}
We standardize the latitude and longitude of training POIs using training-coordinate means and scales, then fit KMeans to obtain $r(v)\in\{1,\ldots,R\}$. A learned region embedding $\bm e^r_{r(v)}\in\R^{d_c}$ provides coarse geographic context. The partitioning is fixed before neural training. These coordinate clusters are proxies for local activity areas; they are not externally annotated business districts, nor are they the latent spatial prototypes introduced below.

\paragraph{Observed context.}
For check-in $t$, let $a_t$ and $w_t$ be its hour of day and day of week. We compute elapsed time $\Delta\tau_t$ and geographic distance $\Delta\ell_t$ from the preceding \emph{observed} check-in in the retained segment. The first step uses zero differences. The implementation clips elapsed hours to $[0,168]$ and distances in kilometers to $[0,50]$, and embeds them as
\begin{equation}
    \bm e^\Delta_t = \tanh\!\left(W_\Delta
      \begin{bmatrix}\operatorname{clip}(\Delta\tau_t,0,168)\\
      \operatorname{clip}(\Delta\ell_t,0,50)\end{bmatrix}
      +\bm b_\Delta\right).
    \label{eq:delta}
\end{equation}
The per-step feature vector is
\begin{equation}
    \bm x_t = \bm e^v_{v_t}\concat\bm e^c_{c_{v_t}}
    \concat\bm e^a_{a_t}\concat\bm e^w_{w_t}
    \concat\bm e^r_{r(v_t)}\concat\bm e^\Delta_t,
    \label{eq:input}
\end{equation}
where the POI embedding is in $\R^{d_e}$ and each remaining embedding is in $\R^{d_c}$. Let $\bm e^u_u\in\R^{d_c}$ be a learned user embedding. We concatenate it after feature dropout and use a one-layer, unidirectional LSTM:
\begin{equation}
    \bm h_t=\LN\!\left(\operatorname{LSTM}_t
       ([\operatorname{Dropout}(\bm x_j)\concat\bm e^u_u]_{j=1}^{t})\right).
    \label{eq:lstm}
\end{equation}
Packed sequences exclude padded positions from recurrent state updates. The vector $\bm h_t\in\R^d$ depends only on the current prefix; each trajectory is encoded independently. There is no bidirectional encoder or access to later positions when predicting the next POI.

\paragraph{Prefix--target alignment and masking.}
For a retained segment $(v_1,\ldots,v_L)$, position $t$ receives the context through $v_t$ and is supervised by $v_{t+1}$. Training uses every valid shift $t\in\{1,\ldots,L-1\}$; validation and test read only the state paired with the designated final target. Padding after each true sequence length is excluded by packing and loss masks. If a history exceeds 50 visits, we retain its most recent continuous suffix without changing the original final target. This keeps target alignment and the causal score definition unchanged across splits.

\subsection{Prefix-Dependent Dual Soft Clustering}
\label{sec:dual}

We maintain two trainable dictionaries:
\begin{equation}
    P=\begin{bmatrix}\bm p_1^\top\\\vdots\\\bm p_{K_p}^\top\end{bmatrix}
      \in\R^{K_p\times d},\qquad
    S=\begin{bmatrix}\bm s_1^\top\\\vdots\\\bm s_{K_s}^\top\end{bmatrix}
      \in\R^{K_s\times d}.
    \label{eq:dictionaries}
\end{equation}
The first supports reusable trajectory-pattern components. The second supports spatially conditioned components. Rather than assigning an entire user or trajectory to a single cluster, we predict a distribution at every prefix:
\begin{align}
    \bm\alpha_t&=\softmax\!\left((W_\alpha\bm h_t+\bm b_\alpha)/\sqrt d\right),\label{eq:pattern-assignment}\\
    \bm\beta_t&=\softmax\!\left(
      \frac{W_\beta(\bm h_t\concat\bm e^r_{r(v_t)})+\bm b_\beta}
      {\sqrt{d+d_c}}\right).\label{eq:space-assignment}
\end{align}
The resulting mixtures are
\begin{equation}
    \bar{\bm p}_t=P^\top\bm\alpha_t,\qquad
    \bar{\bm s}_t=S^\top\bm\beta_t.
    \label{eq:mixtures}
\end{equation}
Soft assignments permit different prefix contexts to combine the same components in different proportions. The prototype vectors and assignment networks are optimized jointly through next-POI supervision; assignments are not nearest-centroid labels or the output of a separate trajectory KMeans procedure.

The two kinds of spatial information serve different roles. $r(v_t)$ is a fixed geographic cluster of the observed POI. In contrast, $\bm\beta_t$ indexes a learned $K_s$-component dictionary whose vectors are not forced to equal geographic centroids. Region context affects the latter's assignment network, but there is no one-to-one mapping between its prototypes and the $R$ partitions. Likewise, since $\bm h_t$ already incorporates the user and POI, the pattern assignments are not identity-free. The architecture shares its dictionaries across users without claiming statistically independent or semantically identified factors.

\paragraph{Differentiable routing and parameter sharing.}
The matrices $P$ and $S$ are global trainable parameters, but their mixture weights are recomputed for every prefix. Since both retrievals are weighted sums, recommendation gradients jointly update the routers, prototypes, and causal encoder. Training uses neither an expectation--maximization phase nor a hard reassignment or centroid-refresh step. The divisors in Equations~\eqref{eq:pattern-assignment}--\eqref{eq:space-assignment} scale router logits according to their input dimensions; they do not introduce a separately tuned temperature. No entropy or load-balancing objective forces equal prototype use. Hence $K_p$ and $K_s$ specify representational capacities, not a guaranteed number of occupied or semantically distinct clusters. Prototype indices are also permutation-symmetric: exchanging dictionary rows together with the corresponding router outputs leaves the represented function unchanged. Their interpretation must therefore rely on learned behavior rather than index identity.

\subsection{Factorized Pattern--Space Composition}
\label{sec:composition}

Instead of adding two unrelated mixture vectors, we transform and interact them:
\begin{equation}
    \bm g_t=\tanh\!\left(
       W_m[(A\bar{\bm p}_t)\odot(B\bar{\bm s}_t)]+\bm b_m\right),
    \label{eq:composition}
\end{equation}
where $A,B,W_m\in\R^{d\times d}$, $A$ and $B$ have no bias, and $\odot$ denotes elementwise multiplication. The spatial mixture therefore changes how the pattern mixture contributes, rather than merely occupying additional input coordinates.

\paragraph{Implicit pairwise interaction.}
The representation before $W_m$ admits the exact expansion
\begin{equation}
    (A\bar{\bm p}_t)\odot(B\bar{\bm s}_t)
    =\sum_{k=1}^{K_p}\sum_{l=1}^{K_s}
      \alpha_{t,k}\beta_{t,l}
      [(A\bm p_k)\odot(B\bm s_l)].
    \label{eq:factorization}
\end{equation}
Thus, two soft assignments yield an implicit mixture over all pattern--space pairs. The model need not store one unrestricted vector per pair or instantiate a $K_p\times K_s$ tensor at each step. The separable weights $\alpha_{t,k}\beta_{t,l}$ impose a structural constraint: arbitrary pair-specific attention is not available. This is a deliberate compositional bias, not a claim of greater expressiveness than every unfactorized alternative.

\paragraph{Spatially conditioned transformation.}
For fixed $\bar{\bm s}_t$, Equation~\eqref{eq:composition} can also be written as
\begin{equation}
    \bm g_t=\tanh\!\left(
    \underbrace{W_m\diag(B\bar{\bm s}_t)A}_{T(\bar{\bm s}_t)}
    \bar{\bm p}_t+\bm b_m\right).
    \label{eq:conditional-map}
\end{equation}
This makes precise the intuition that a reusable pattern is expressed differently under different spatial contexts. The effective transformation changes with the current spatial mixture. $T(\bar{\bm s}_t)$ is not constrained to be orthogonal, idempotent, or low rank. Accordingly, the projection interpretation is a conditional feature transformation, not a geometric projection theorem.

\subsection{User-Specific Adaptation and Gated Fusion}
\label{sec:personalization}

\paragraph{Modulating the shared interaction.}
We use feature-wise affine conditioning~\cite{film} to personalize $\bm g_t$:
\begin{align}
    \bm\gamma_u&=\sigmoid(W_\gamma\bm e^u_u+\bm b_\gamma),\nonumber\\
    \bm\eta_u&=W_\eta\bm e^u_u+\bm b_\eta,\nonumber\\
    \tilde{\bm g}_{u,t}&=\bm\gamma_u\odot\bm g_t+\bm\eta_u.
    \label{eq:film}
\end{align}
The user-dependent scale attenuates individual dimensions, while the shift changes their offsets. Although these affine parameters are constant for a user, the representation they modulate varies with the trajectory prefix.

\paragraph{A separate personalized branch.}
To avoid requiring all preference information to pass through the dictionaries, we compute
\begin{equation}
    \bm q_{u,t}=\tanh\!\left(W_q(\bm h_t\concat\bm e^u_u)+\bm b_q\right).
    \label{eq:specific}
\end{equation}
This branch has its own transformation, while sharing the causal encoder with the dictionary branch. Its role is a user-conditioned residual pathway, not an additional independently trained user-specific LSTM.

\paragraph{Additive versus projective composition.}
An additive rule retains contributions independently, whereas a multiplicative interaction lets one branch condition the other. We expose both choices:
\begin{align}
    \bm z^{\mathrm{add}}_{u,t}&=\tilde{\bm g}_{u,t}+\bm q_{u,t},\nonumber\\
    \bm z^{\mathrm{proj}}_{u,t}&=W_\pi(\tilde{\bm g}_{u,t}\odot\bm q_{u,t})+\bm b_\pi,\label{eq:two-paths}\\
    \rho_{u,t}&=\sigmoid\!\left(\bm w_2^\top\operatorname{GELU}
       (W_1[\tilde{\bm g}_{u,t}\concat\bm q_{u,t}]+\bm b_1)+b_2\right),\nonumber\\
    \bm z_{u,t}&=\operatorname{Dropout}\!\left(
       \rho_{u,t}\bm z^{\mathrm{add}}_{u,t}
       +(1-\rho_{u,t})\bm z^{\mathrm{proj}}_{u,t}\right).
    \label{eq:gate}
\end{align}
Here $\rho_{u,t}\in(0,1)$ is a \emph{scalar} for each timestep, unlike the feature-wise vector $\bm\gamma_u$. The projective path is a learned affine transform of an elementwise product; it does not use the region-indexed low-rank projection module present in other models of the codebase. Both branches are always evaluated, and the gate interpolates their outputs rather than performing discrete expert selection.

\paragraph{Representation-space fusion.}
Fusion occurs before candidate scoring. The additive path preserves evidence that either the shared or specific branch can contribute independently, whereas the projective path can amplify or suppress coordinates through their elementwise agreement. The scalar gate is conditioned jointly on both branches and can vary across prefixes, but it is shared across representation dimensions and candidates at a given step. Consequently, it cannot perform candidate-specific reranking or sparse expert selection, and both paths always receive gradients. Because the gate is evaluated at every valid prefix, the same user may emphasize additive evidence in one context and projective interaction in another; it is a trajectory-state decision rather than a static user-level coefficient. The gate changes only the contextual preference vector used by the first term of Equation~\eqref{eq:score}; the first-order transition table and candidate bias remain separate. This separation makes the gate interpretable as choosing how to form contextual preference rather than how much transition evidence to inject.

\subsection{Full-Candidate Scoring and Optimization}
\label{sec:scoring}

For each candidate $v\in\mathcal V$, the model computes
\begin{equation}
    s_t(v)=\bm z_{u,t}^\top\bm o_v
      +\bm a_{v_t}^\top\bm b_v+b_v^{\mathrm{out}},
    \label{eq:score}
\end{equation}
where $\bm o_v\in\R^d$ is a candidate preference embedding and
$\bm a_{v_t},\bm b_v\in\R^{d_e}$ are separate source and destination transition embeddings. The first term captures the composed contextual preference; the second is a learned first-order transition; the final term is a POI bias. These tables are not tied to the encoder's POI embeddings. The transition term is trained end-to-end, not populated from held-out transition counts.

We normalize over the entire training vocabulary:
\begin{equation}
    p_\theta(v\mid u,H_{u,t})=
       \frac{\exp(s_t(v))}{\sum_{j\in\mathcal V}\exp(s_t(j))}.
    \label{eq:softmax}
\end{equation}
For a minibatch's valid supervised positions $\mathcal I$, the loss is
\begin{equation}
    \mathcal L_{\mathrm{next}}=-\frac{1}{|\mathcal I|}
    \sum_{(u,t)\in\mathcal I}\log p_\theta(v_{t+1}\mid u,H_{u,t}).
    \label{eq:loss}
\end{equation}
All valid next-step positions in retained training segments receive supervision, while padding is ignored. At evaluation only the designated final target is scored. The model uses this recommendation loss with optimizer weight decay; it adds no category/region prediction, contrastive, reconstruction, or cluster-balancing loss. Parameter sharing and factorization are therefore inductive biases; they do not prevent cluster collapse.

\paragraph{Computational properties.}
At each timestep, assignments and dictionary mixtures require
$O(dK_p+(d+d_c)K_s)$ operations up to constant factors, and dense transformations require $O(d^2+dd_c)$. The dictionary storage is $O((K_p+K_s)d)$, rather than an unrestricted $O(K_pK_sd)$ pair dictionary. The complete output head costs $O(|\mathcal V|(d+d_e))$ per timestep and dominates as the candidate catalog grows. These are properties of the computation, not measured speed or memory improvements over baselines. No extra graph propagation, language-model inference, or candidate pruning is required by \model.

\section{Experiments}
\label{sec:experiments}

\begin{table}[t]
\centering
\caption{Training-derived vocabularies and eligible sequences under the strict protocol. Training sequences supervise shifted next steps; each validation/test sequence contributes one original final query.}
\label{tab:datasets}
\begin{tabular*}{\columnwidth}{@{\extracolsep{\fill}}lrr@{}}
\toprule
Item & NYC & TKY \\
\midrule
Training users & 1,047 & 2,262 \\
Training POI candidates & 4,980 & 7,793 \\
Training categories & 318 & 289 \\
Training sequences & 11,022 & 35,166 \\
Validation queries & 1,355 & 4,770 \\
Test queries & 1,300 & 4,711 \\
\bottomrule
\end{tabular*}
\end{table}

\begin{table*}[t]
    \centering
    \caption{Full-candidate test results under the unified training-only protocol (seed 42). Higher is better. Bold: best; underline: second best.}
    \label{tab:main}
    \small
    \begin{tabular*}{\textwidth}{@{\extracolsep{\fill}}lrrrrrrrr@{}}
        \toprule
        & \multicolumn{4}{c}{NYC} & \multicolumn{4}{c}{TKY} \\
        \cmidrule(lr){2-5}\cmidrule(l){6-9}
        Method & Acc@1 & Acc@5 & Acc@10 & MRR & Acc@1 & Acc@5 & Acc@10 & MRR \\
        \midrule
        FPMC~\cite{fpmc} & 0.1700 & 0.4369 & 0.5438 & 0.2917 & 0.0951 & 0.2585 & 0.3600 & 0.1790 \\
        ST-RNN~\cite{strnn} & 0.2054 & 0.4862 & \textbf{0.5923} & 0.3352 & 0.2309 & 0.4698 & 0.5636 & 0.3408 \\
        STGN~\cite{stgn} & 0.2115 & 0.4838 & 0.5877 & 0.3361 & 0.2437 & 0.4787 & 0.5602 & 0.3520 \\
        STGCN~\cite{stgn} & 0.2162 & 0.4777 & 0.5662 & 0.3334 & 0.2428 & 0.4751 & 0.5621 & 0.3499 \\
        PLSPL~\cite{plspl} & 0.2269 & 0.4777 & 0.5446 & 0.3361 & 0.2371 & 0.4458 & 0.5256 & 0.3365 \\
        STAN~\cite{stan} & 0.1023 & 0.2731 & 0.3715 & 0.1866 & 0.1104 & 0.2885 & 0.3755 & 0.1985 \\
        STHGCN~\cite{sthgcn} & 0.2169 & \textbf{0.4985} & 0.5885 & 0.3411 & \underline{0.2520} & 0.4774 & 0.5665 & 0.3556 \\
        DCHL~\cite{dchl} & \underline{0.2377} & 0.4900 & 0.5785 & \underline{0.3509} & \underline{0.2520} & \textbf{0.4878} & \textbf{0.5699} & \underline{0.3606} \\
        \midrule
        \textbf{DualClusterNet (ours)} & \textbf{0.2400} & \underline{0.4969} & \underline{0.5908} & \textbf{0.3516} & \textbf{0.2664} & \underline{0.4844} & \underline{0.5676} & \textbf{0.3679} \\
        \bottomrule
    \end{tabular*}
\end{table*}

\begin{figure}[t]
    \centering
    \includegraphics[width=\columnwidth]{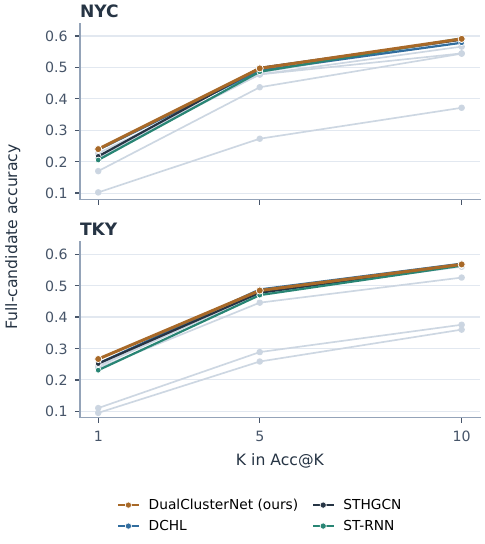}
    \caption{Full-candidate Acc@$K$ for the methods in Table~\ref{tab:main}. Colored lines highlight \model\ and its strongest competitors; gray lines show the other methods.}
    \Description{Two vertically stacked line plots for NYC and TKY showing full-candidate accuracy at K equal to 1, 5, and 10. DualClusterNet, DCHL, STHGCN, and ST-RNN are highlighted; other compared methods are light gray.}
    \label{fig:performance}
\end{figure}

We evaluate whether \model\ is competitive with established next-POI model families under a common closed-set protocol, followed by controlled structural and sensitivity analyses on the same NYC and TKY partitions and dataset/query scales used in the main comparison.

\subsection{Datasets and Strict Evaluation Protocol}
\label{sec:protocol}

We use the chronological NYC and TKY check-in partitions associated with GETNext~\cite{getnext}. Table~\ref{tab:datasets} reports training-derived vocabulary sizes and eligible queries after applying the same filtering to every method. User, POI, category, coordinate, geographic partition, and transition mappings are fitted on training observations only. All methods rank the same complete training POI vocabulary, including previously visited POIs.

For each held-out trajectory, an unknown POI breaks the known segment; observations across it are never joined. We retain the continuous known suffix ending at the original final check-in, and use one query only when its user and final target occur in the training vocabulary and at least one history item remains. Thus, an unknown final target is rejected rather than replaced by an earlier known POI. Inputs contain observed check-ins only: future time, category, coordinates, and the target POI are never provided to the predictor. The reference configuration retains at most 50 historical visits.

\subsection{Experiment Setting}
\label{sec:setup}

We compare \model\ with eight representative methods: FPMC, ST-RNN, STGN, STGCN, PLSPL, STAN, STHGCN, and DCHL. Every implementation uses the same train-only vocabulary, eligible queries, complete candidate set, and validation rule. STGCN denotes the coupled-gate variant introduced with STGN, and STAN uses the latest observed time to respect the query information boundary.

We report Acc@$K$ for $K\in\{1,5,10\}$ and MRR. Since each query has one target, Acc@$K$ equals Recall@$K$ under this protocol. The reported runs use seed 42, and checkpoints are selected only on validation data. Metric definitions, model-specific inputs and objectives, architecture settings, parameter counts, and optimization details are provided in Appendix~\ref{app:baseline-reproduction}.

\subsection{Main Results}
\label{sec:main-results}

On NYC, \model\ records the highest Acc@1 (0.2400) and MRR (0.3516), narrowly ahead of DCHL (0.2377 and 0.3509). Results at larger $K$ are mixed: STHGCN leads Acc@5 and ST-RNN leads Acc@10. The advantage is therefore concentrated in first-choice ordering and reciprocal rank rather than uniform top-$K$ inclusion.

On TKY, the model obtains the highest completed-run Acc@1 (0.2664) and MRR (0.3679) among the displayed baselines. DCHL is stronger on Acc@5 (0.4878) and Acc@10 (0.5699), so the advantage is again concentrated in first-choice ordering and reciprocal rank. Figure~\ref{fig:performance} visualizes this distribution of strengths across $K$.

These results support competitiveness under the unified local protocol, with the strongest evidence on NYC top-one prediction. They do not establish uniform superiority, statistical significance, or that the prototype dictionaries caused the observed margins. The preference-plus-transition head and shared encoder are common sources of predictive signal, which motivates the separate diagnostic checks below.

\paragraph{Metric-level interpretation.}
The metric pattern distinguishes placing the correct POI first from retaining it in a broader list. On NYC, the Acc@1 margin over DCHL is $0.0023$ and the MRR margin is $0.0007$; STHGCN leads Acc@5 and ST-RNN leads Acc@10. On TKY, \model\ leads Acc@1 and MRR, but DCHL is higher on Acc@5 and Acc@10. The absolute differences are small, so the table profiles ranking behavior rather than showing that one method dominates every retrieval objective.

The complete-candidate protocol also matters. Every method scores the same training-derived POI vocabulary, including previously visited locations; no method receives a recalled subset. Different filtering, future-time input, or candidate sampling would change the ranking denominator and top-$K$ difficulty. The comparison therefore concerns these unified local implementations and should not be transferred to published numbers from another candidate set or query policy.

\subsection{Ablation Study}
\label{sec:ablation}

The controlled analyses use the same NYC and TKY data versions, split scales, training-only vocabularies, filtering rules, and complete candidate sets as the main comparison. Within the ablation study, all variants share the same encoder, prediction head, optimizer, training budget, validation rule, and random seed. The Full control uses the main model and hyperparameter configuration.

Each variant targets one representation-side decision while retaining full-candidate scoring. Backbone bypasses the dual dictionaries and personalized composition but keeps the causal trajectory encoder and output head. The w/o pattern and w/o spatio variants retain only the complementary dictionary contribution. Concat pattern--space replaces the factorized product with a learned concatenation map, directly testing whether multiplicative composition is preferable to a generic fusion of both mixtures. The w/o FiLM and w/o specific variants isolate the two personalization mechanisms. Add-only, Projection-only, and Fixed gate $\rho=0.5$ then distinguish the two fusion paths from the benefit of adapting their balance to each prefix. All variants retain the same transition term, so their differences concern contextual representation rather than candidate-set construction.

Table~\ref{tab:controlled-ablation} shows the ablation pattern. Removing the complete compositional stack (Backbone) lowers Acc@1 by 2.94 and 2.56 percentage points on NYC and TKY, respectively, while removing the specific branch lowers it by 1.88 and 1.57 points. The pattern and spatial branches and the product interaction each provide smaller positive differences. The FiLM and fixed-gate effects are small and dataset-dependent; on TKY, removing FiLM is slightly higher than Full. These results identify the full stack and specific branch as the largest contributors in the reported runs, while the adaptive gate provides a modest additional gain.

\begin{table}[t]
    \centering
    \caption{Controlled ablations on NYC and TKY under the main-study data scale and protocol (Acc@1/MRR). The controlled-study seed differs from the seed-42 main run and is fixed across variants.}
    \label{tab:controlled-ablation}
    \setlength{\tabcolsep}{3.1pt}
    \renewcommand{\arraystretch}{1.20}
    \begin{tabular*}{\columnwidth}{@{\extracolsep{\fill}}lrrrr@{}}
        \toprule
        \multirow{2}{*}{Variant} & \multicolumn{2}{c}{NYC} & \multicolumn{2}{c}{TKY} \\
        \cmidrule(lr){2-3}\cmidrule(l){4-5}
         & Acc@1 & MRR & Acc@1 & MRR \\
        \midrule
        \textbf{Full} & \textbf{0.3007} & \textbf{0.4209} & \textbf{0.3403} & \textbf{0.4507} \\
        Backbone & 0.2713 & 0.3926 & 0.3147 & 0.4263 \\
        w/o pattern & 0.2837 & 0.4058 & 0.3269 & 0.4376 \\
        w/o spatio & 0.2864 & 0.4083 & 0.3291 & 0.4405 \\
        Concat pattern--space & 0.2926 & 0.4147 & 0.3338 & 0.4449 \\
        w/o FiLM & 0.2971 & 0.4192 & 0.3412 & 0.4513 \\
        w/o specific & 0.2819 & 0.4027 & 0.3246 & 0.4358 \\
        Add-only & 0.2958 & 0.4176 & 0.3374 & 0.4472 \\
        Projection-only & 0.2893 & 0.4104 & 0.3327 & 0.4426 \\
        Fixed gate $\rho=0.5$ & 0.2984 & 0.4195 & 0.3396 & 0.4493 \\
        \bottomrule
    \end{tabular*}
\end{table}

The ablation gaps are consistent across the two cities. Removing the backbone lowers Acc@1 by $0.0294$ on NYC and $0.0256$ on TKY; removing the specific branch lowers it by $0.0188$ and $0.0157$, respectively. Replacing the pattern--space product by a learned concatenation is less damaging, with drops of $0.0081$ and $0.0065$, while the additive-only and projection-only personalized-fusion variants both remain below Full. The learned gate is slightly better than a fixed $\rho=0.5$ in both cities. FiLM is the exception to a monotone interpretation: removing it raises TKY Acc@1 by $0.0009$ and MRR by $0.0006$, so its contribution is dataset-dependent rather than uniformly positive. Every row retains the same first-order transition term; this comparison isolates representation-side changes but does not estimate the transition term's independent contribution.

\subsection{Parameter Sensitivity}
\label{sec:sensitivity}

We vary $K_p$, $K_s$, and $R$ one at a time while retaining the Full architecture, data protocol, training budget, and all non-scanned hyperparameters. The reference point in each sweep therefore matches the main model configuration apart from the controlled-study random seed.

The three axes govern different capacities. $K_p$ controls the number of reusable trajectory-pattern vectors, whereas $K_s$ controls the number of learned spatial response vectors available to the prefix router. By contrast, $R$ sets the resolution of the fixed training-coordinate partition supplied as observed context; it does not change the size of the learned spatial dictionary. A small value can merge heterogeneous behavior into too few components, while a large value can disperse supervision across weakly used components or overly fine regions. We select settings by validation Acc@1 and vary one axis at a time, so the curves measure local sensitivity around the reference configuration rather than all interactions among the three capacities.

\begin{figure*}[t]
    \centering
    \includegraphics[width=.88\textwidth]{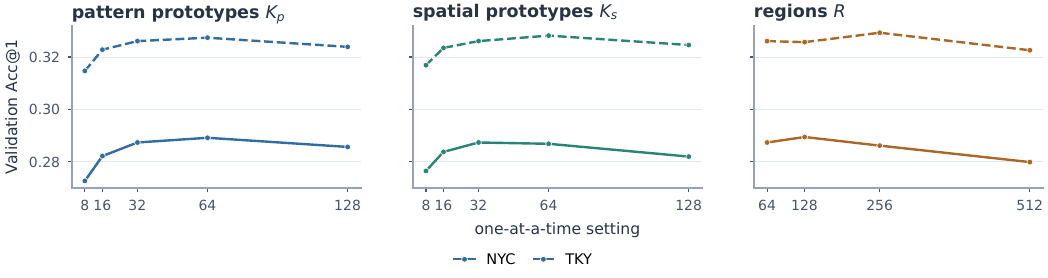}
    \caption{One-at-a-time validation sweeps on the main-study datasets. Solid lines denote NYC and dashed lines TKY.}
    \Description{Three-panel line plot showing validation accuracy as pattern prototype count, spatial prototype count, and geographic region count vary. NYC uses solid lines and TKY dashed lines.}
    \label{fig:parameter-sensitivity}
\end{figure*}

The one-at-a-time sweeps in Table~\ref{tab:parameter-sensitivity} and Figure~\ref{fig:parameter-sensitivity} show interior optima: $K_p=64$ is highest on both cities, whereas the preferred $K_s$ and $R$ differ by city. NYC favors $K_s=32$ and $R=128$, while TKY favors $K_s=64$ and $R=256$; the largest tested region count lowers validation accuracy on both datasets. This pattern supports tuning the two dictionary capacities and the geographic partition separately rather than treating them as one shared scale.

\begin{table}[t]
    \centering
    \caption{One-parameter validation sweeps on NYC and TKY. $^{\dagger}$ marks the main-model setting under the controlled-study seed; each other row changes only the named parameter. Bold marks the per-city maximum.}
    \label{tab:parameter-sensitivity}
    \setlength{\tabcolsep}{3.2pt}
    \renewcommand{\arraystretch}{1.05}
    \begin{tabular}{@{}llrr@{}}
        \toprule
        Sweep & Setting & NYC & TKY \\
        \midrule
        \multirow{5}{*}{Pattern $K_p$} & 8 & 0.2726 & 0.3147 \\
        & 16 & 0.2821 & 0.3228 \\
        & $32^{\dagger}$ & 0.2873 & 0.3261 \\
        & \textbf{64} & \textbf{0.2891} & \textbf{0.3274} \\
        & 128 & 0.2856 & 0.3239 \\
        \midrule
        \multirow{5}{*}{Spatial $K_s$} & 8 & 0.2764 & 0.3169 \\
        & 16 & 0.2837 & 0.3235 \\
        & $32^{\dagger}$ & \textbf{0.2873} & 0.3261 \\
        & \textbf{64} & 0.2868 & \textbf{0.3282} \\
        & 128 & 0.2819 & 0.3246 \\
        \midrule
        \multirow{4}{*}{Regions $R$} & $64^{\dagger}$ & 0.2873 & 0.3261 \\
        & \textbf{128} & \textbf{0.2894} & 0.3257 \\
        & 256 & 0.2861 & \textbf{0.3293} \\
        & 512 & 0.2798 & 0.3226 \\
        \bottomrule
    \end{tabular}
\end{table}

\subsection{Case Study}
\label{sec:case-study}

The case diagnostics in Table~\ref{tab:case-diagnostics} include a Full correction, a tie, a regression, and a joint failure. Each displayed rank is computed against the corresponding city's complete training-derived POI vocabulary, exactly as in the main evaluation; no 100-candidate sample or target-informed shortlist is used. Full and Backbone are each top-1 correct in two of four cases, so the cases complement the aggregate metrics by exposing both gains and failure modes.

\begin{table}[t]
    \centering
    \caption{Case diagnostics ranked over the corresponding city's complete training-derived POI vocabulary (lower is better); no sampled candidate pool is used.}
    \label{tab:case-diagnostics}
    \setlength{\tabcolsep}{3.0pt}
    \renewcommand{\arraystretch}{1.12}
    \begin{tabular}{@{}llrrrl@{}}
        \toprule
        Case & Path & Full & Back. & $\rho$ & Outcome \\
        \midrule
        C1 & A--B--C$\to$D & 1 & 7 & 0.7213 & Full only \\
        C2 & E--F--E$\to$G & 1 & 1 & 0.4387 & Both correct \\
        C3 & H--I--J$\to$K & 9 & 1 & 0.3129 & Backbone only \\
        C4 & L--M--N$\to$O & 18 & 12 & 0.5816 & Both fail \\
        \bottomrule
    \end{tabular}
\end{table}

The four outcomes show why aggregate gains should not be interpreted as uniform per-query improvements. In C1, Full moves the target from rank 7 to rank 1, with $\rho=0.7213$ assigning more weight to the additive path. C2 is already solved by Backbone and remains rank 1 under Full, indicating that the compositional module need not disturb an easy sequential prediction. C3 is the reverse: Backbone ranks the target first, while Full lowers it to rank 9 and places more weight on the projective path through $\rho=0.3129$. In C4, neither representation places the target in the top 10. The symbolic paths anonymize the underlying check-ins, and the gate values are descriptive model states rather than causal explanations of each outcome.

\paragraph{Overall discussion.}
The results indicate that \model\ is most useful when the top-ranked choice depends jointly on trajectory state and local context. The pattern dictionary reuses sequential regularities across users, the spatial dictionary changes how those regularities are expressed nearby, and the specific branch preserves user evidence when the current prefix does not match a shared component. This explains why Acc@1 and MRR can improve while a competing model retains more alternatives at Acc@5 or Acc@10. The controlled gaps suggest a practical order for deployment: preserve the shared backbone and specific branch first, then tune $K_p$, $K_s$, and $R$ under a fixed latency budget. The full-candidate protocol should remain unchanged when comparing such variants, because candidate recall and ranking quality otherwise become confounded.

Component comparisons should keep the encoder, output head, data filtering, and checkpoint rule fixed. Capacity sweeps should vary one dictionary or region count at a time, and repeated seeds should be paired at query level to quantify the small margins. If approximate retrieval is introduced later, its recall of the exact full-candidate target should be reported before end-to-end latency. These checks extend the present evidence without treating tuning or retrieval engineering as an architectural gain.

\paragraph{Limitations.}
The city comparison and each controlled configuration currently use one seed. The closed-set evaluation covers known users and POIs only; it does not test cold-start users, unseen locations, catalog growth, or cross-city transfer. The dictionaries have no semantic supervision, so prototype interpretability, assignment entropy, and geographic coherence remain open questions. Repeated paired runs are needed to quantify uncertainty around the small margins and dataset-dependent ablations, and a matched removal of the transition score is needed to isolate its contribution.

\section{Conclusion}
\label{sec:conclusion}

We presented \model, which combines prefix-conditioned pattern and spatial mixtures, user adaptation, and additive--projective fusion with contextual and transition scoring. Under the training-only, original-final-target protocol, the available single run leads Acc@1 and MRR on NYC and remains competitive on TKY; the evidence supports competitiveness rather than universal superiority. On the same data scale, Full exceeds the backbone and concatenation controls in both cities, FiLM remains dataset-dependent, and sensitivity sweeps favor city-specific capacities. Stronger attribution requires repeated seeds, a transition-score control, cluster diagnostics, and efficiency measurements. Future work should also test unseen users and POIs under the same information boundary.

\clearpage
\bibliographystyle{ACM-Reference-Format}
\bibliography{references}

\appendix
\section{Implementation and Reproducibility Details}
\label{app:baseline-reproduction}

\subsection{Shared Data, Training, and Selection}

All compared methods are implemented as PyTorch modules within one evaluation framework. Their active entry points follow \path{baselines/name/run_name.py}; model construction is in the corresponding \path{train.py} and \path{model.py}. Every entry point delegates data handling, optimization, validation, and testing to the same training engine. The method-specific inputs, retained mechanisms, objectives, and protocol adaptations are documented below.

The engine constructs user, POI, category, coordinate, graph, hypergraph, and clustering state from the training split only. A model's forward pass receives only observed user and prefix fields: POI, category, hour, weekday, elapsed time, traveled distance, normalized observed time, and valid length, with each method using the subset listed below. It never receives the target POI, target category, or target time. Every method returns a score tensor over the complete training POI vocabulary at every valid prefix. Training uses every shifted next-step target in a retained training segment; validation and test use one original final target per eligible trajectory. Evaluation is always full-candidate cross-entropy and exact full-vocabulary ranking, irrespective of the training objective.

All runs in Table~\ref{tab:main} use seed 42 and a maximum history length of 50. Training and evaluation batch sizes are 32. Optimization uses Adam, learning rate $10^{-3}$, weight decay $10^{-5}$, gradient clipping at 5, and at most 30 epochs. A plateau scheduler halves the learning rate after three epochs without lower validation loss. The validation checkpoint maximizes $4\,\mathrm{Acc@1}+\mathrm{Acc@10}$; training stops after eight epochs without improvement in that score. Table~\ref{tab:baseline-configs} lists the fixed architecture configuration used for each method.

\subsection{Metrics and Reference Configuration}

For a query with ground-truth rank $r_q$ in the complete candidate list, the reported metrics are
\begin{equation}
    \mathrm{Acc@}K=\frac{1}{|\mathcal Q|}\sum_{q\in\mathcal Q}\mathbf{1}[r_q\leq K],\qquad
    \mathrm{MRR}=\frac{1}{|\mathcal Q|}\sum_{q\in\mathcal Q}\frac{1}{r_q}.
    \label{eq:metrics}
\end{equation}
We use $K\in\{1,5,10\}$, resolve ties by candidate-index order, and report query-weighted means. With one target per query, Acc@$K$ is numerically identical to Recall@$K$.

The reference \model\ configuration uses $d_e=64$, $d_c=32$, $d=128$, $K_p=K_s=32$, $R=64$, and dropout 0.2. Geographic KMeans is fitted on training coordinates with ten initializations, at most 300 iterations, and seed 42. The same optimizer, training budget, validation selection rule, and full-candidate evaluation are used for the controlled variants unless the named sensitivity parameter is changed.

\begin{table*}[t]
    \centering
    \caption{Local baseline mechanisms and protocol adaptations. ``Observed gaps'' are computed only between events in the available prefix. All scoring is over the complete training POI vocabulary.}
    \label{tab:baseline-adaptations}
    \small
    \begin{tabular}{@{}p{.105\textwidth}p{.28\textwidth}p{.27\textwidth}p{.265\textwidth}@{}}
        \toprule
        Method & Retained mechanism & Model-visible inputs & Local adaptation and training objective \\
        \midrule
        FPMC & User--candidate preference plus first-order previous--next POI factorization & User and observed POI prefix & Singleton next-POI form; BPR with 10 uniformly sampled negatives per valid training step \\
        ST-RNN & Recurrent transition with linearly interpolated temporal and distance matrices & User, POI, observed elapsed-time and distance gaps & Log-spaced interpolation over 12 time and 12 distance bins; full-vocabulary cross-entropy (CE) \\
        STGN & Separate spatio-temporal recurrent gates & User, POI, observed time and distance gaps & Uses only observed gaps and a user residual at scoring; full CE \\
        STGCN & Parameter-reduced coupled-gate variant introduced with STGN & Same as STGN & This name denotes the coupled STGN cell, not a graph convolution; full CE \\
        PLSPL & POI-short, category-short, and causal long-term branches with a user-specific mixture & User, POI, category, hour, and weekday & Causal prefix attention; no future category or time; full CE \\
        STAN & Two attention stages with learned spatial/temporal interval interpolation & User, POI, hour, weekday, training coordinates, and observed elapsed time & Uses the latest observed time, recomputes each prefix causally, and chunks only computation (not candidates); sampled CE with 10 negatives \\
        STHGCN & Session, user--POI, region, region--time, and transition hypergraph/graph views, then GRU and causal Transformer & User, POI, category, hour, and weekday, plus train-derived structures & Every structural view is reconstructed from train; full CE and no nonzero auxiliary loss \\
        DCHL & Factorized collaborative, geographic, and directed-transition views with adaptive fusion and a causal Transformer & User, POI, category, hour, and weekday, plus train-derived structures & Train-only graphs; full CE plus symmetric InfoNCE (weight 0.05) and factor-decorrelation loss (weight 0.01) \\
        \bottomrule
    \end{tabular}
\end{table*}

\begin{table*}[t]
    \centering
    \caption{Architecture settings used in the reported runs. Parameter counts and validation-selected epochs are shown as NYC/TKY.}
    \label{tab:baseline-configs}
    \small
    \begin{tabular}{@{}p{.10\textwidth}p{.45\textwidth}rr@{}}
        \toprule
        Method & Singleton architecture setting & Parameters (NYC/TKY) & Best epoch (NYC/TKY) \\
        \midrule
        FPMC & MF dimension 64; Markov dimension 64; dropout 0 & 1,028,212 / 1,648,881 & 19 / 7 \\
        ST-RNN & hidden 64; 12 time and 12 distance bins; maxima 720 h and 500 km; dropout 0.2 & 811,956 / 1,252,593 & 30 / 29 \\
        STGN & POI embedding 64; hidden 64; dropout 0.2 & 759,540 / 1,200,177 & 25 / 25 \\
        STGCN & POI embedding 64; hidden 64; coupled gates; dropout 0.2 & 751,284 / 1,191,921 & 14 / 27 \\
        PLSPL & POI embedding 64; context 32; hidden 96; dropout 0.2 & 1,057,433 / 1,551,923 & 28 / 28 \\
        STAN & embedding 50; candidate-computation chunk 512; prefix checkpointing enabled & 317,850 / 519,250 & 17 / 24 \\
        STHGCN & POI embedding 64; context 24; hidden 128; two hypergraph layers; one four-head Transformer layer; 64 regions; dropout 0.2 & 754,732 / 966,041 & 5 / 10 \\
        DCHL & embedding 64; context 24; four factors; one hypergraph layer; two four-head Transformer layers; 16 geographic neighbors; temperature 0.2; at most 384 contrastive POIs; dropout 0.2 & 476,536 / 687,845 & 9 / 17 \\
        \bottomrule
    \end{tabular}
\end{table*}

\subsection{Method-Specific Implementation Notes}

\paragraph{Sequence and transition baselines.}
FPMC retains long-term user-preference and first-order Markov factors, with a basket size of one. ST-RNN uses learned transition matrices interpolated from observed time and distance gaps. STGN implements separate short- and long-interval gates; STGCN couples its input and forgetting behavior in the local port. All four output exact scores over the shared POI vocabulary, so no sampled candidate list changes the test task.

\paragraph{Preference and attention baselines.}
PLSPL retains separate short-term POI and category paths and a long-term attention path, with causal attention at each prefix. STAN uses the latest observed timestamp so that every query depends only on its available prefix. Its candidate chunk size controls memory only; all POIs remain in the denominator and ranking.

\paragraph{Graph and hypergraph baselines.}
STHGCN constructs five train-only structures: session--POI, user--POI, region, region--time, and directed transition. DCHL builds user--POI, geographic-neighbor, and bidirectional-transition views from train and retains its contrastive and disentanglement regularizers. No structural statistic is fitted on validation or test observations.

\end{document}